\documentclass[Times,twocolumn,tighten,deluxetable]{aastex63}

\usepackage{graphicx}
\usepackage{rotating}
\usepackage{lineno}
\usepackage{multirow}
\usepackage{enumerate}

\submitjournal{ApJ}

\shorttitle{Intrinsic Parameters of MAXI J1834-021}
\shortauthors{Debnath et al.}

\begin{document}

\title{Evolution of QPO During Discovery Outburst of MAXI J1834-021: Estimation of Intrinsic Parameters from Spectro-Temporal Study}

\correspondingauthor{Dipak Debnath}
\email{dipakcsp@gmail.com}

\author[0000-0003-1856-5504]{Dipak Debnath}
\affiliation{Institute of Astronomy, National Tsing Hua University, Hsinchu 300044, Taiwan}
\affiliation{Institute of Astronomy Space and Earth Science, P 177, CIT Road, Scheme 7m, Kolkata 700054, India}

\author[0000-0002-5617-3117]{Hsiang-Kuang Chang}
\affiliation{Institute of Astronomy, National Tsing Hua University, Hsinchu 300044, Taiwan}
\affiliation{Department of Physics, National Tsing Hua University, Hsinchu 300044, Taiwan}



\begin{abstract}
The Galactic transient black hole candidate (BHC) MAXI~J1834-021 was detected for the first time by MAXI/GSC on February 05, 2023 and 
it was active for next $\sim 10$~months. A monotonic evolution of low-frequency QPOs from higher to lower frequencies is observed in 
the middle-phase of the outburst. We study this evolution of the QPO with the propagating oscillatory shock (POS) model, and it suggests 
the presence of a receding shock. The POS model fit also estimates the mass of the source to be $12.1\pm0.3~M_\odot$. We also study the 
boradband ($0.5$-$70$~keV) nature of the source using archival data of NICER and NuSTAR on March 10, 2023 with the both phenomenological 
(combined disk blackbody plus powerlaw) and physical (\textit{nthComp, kerrbb, TCAF}) models. The mass of the BHC estimated by the 
\texttt{kerbb} and \texttt{TCAF} models is found to be consistent with POS model fits as well as reported in our recent work. Combining 
all these methods, we predict mass of the source as $12.3^{+1.1}_{-2.0}~M_\odot$. The \texttt{kerbb} model fit also estimates the spin, 
distance, and inclination of the source to be $0.13^{+0.03}_{-0.02}$, $9.2^{+0.4}_{-0.9}$~kpc, and $80^\circ.0$$^{+2.7}_{-6.0}$, 
respectively. The combined spectral study suggests harder spectral state of the source with a higher dominance of the sub-Keplerian halo 
accretion rate over the Keplerian disk rate. The consistency of the observed frequency of the QPO with that of obtained from the 
\texttt{TCAF} model fitted shock parameters, confirms shock oscillation as the origin of the QPO. 
\end{abstract}


\section{Introduction}

The black hole (BH) continuum spectrum mainly consists of two components: thermal multi-color disk blackbody (DBB) and non-thermal power-law (PL)
components. In addition to ionized line emissions, strong reflection features from ionized plasma are sometimes observed in the BH spectrum. 
To understand physics of accretion processes around the BHs, one needs to perform a detailed study of the spectral and temporal properties of the 
sources using physical accretion disk models. These models provide a clear understanding about the flow dynamics of BH X-ray binaries during their 
active phases. {\fontfamily{qcr}\selectfont nthComp} \citep{Zdziarski96,Zycki99}, {\fontfamily{qcr}\selectfont kerrbb} \citep{Li05}, 
{\fontfamily{qcr}\selectfont pexrav} \citep{Magdziarz95}, {\fontfamily{qcr}\selectfont relxill} \citep[][and references therein]{Garcia14}, 
and {\fontfamily{qcr}\selectfont TCAF} \citep{CT95,D14,D15} are some of the widely used physical models. The {\fontfamily{qcr}\selectfont nthComp} 
model is used to understand nonthermal Comptonized flux contributions and its high-energy roll-over, while the {\fontfamily{qcr}\selectfont kerrbb} 
model is a multi-temperature blackbody model for a thin relativistic accretion disk around a Kerr BH. The {\fontfamily{qcr}\selectfont pexrav} model 
describes an exponentially cut-off power-law spectrum reflected from neutral material, such as Fe and Ni, while {\fontfamily{qcr}\selectfont relxill} 
model is a relativistic accretion disk model that effectively explains the reflection features observed in BHs. The {\fontfamily{qcr}\selectfont TCAF} 
model considers two types of accretion flows: a Keplerian disk (high viscosity, high angular momentum, geometrically thin, and optically thick) and 
a sub-Keplerian halo (low viscosity, low angular momentum, geometrically thick, and optically thin) to explain the physical processes around BHs. 
These physical models also provide reliable estimates of intrinsic source parameters such as mass, spin, distance, and inclination angle.

In general, four spectral states—hard (HS), hard-intermediate (HIMS), soft-intermediate (SIMS), and soft (SS)—are observed during the outburst 
of a transient black hole candidate (BHC). These states are classified based on their distinct spectral and temporal properties 
\citep{RM06,Belloni05,Nandi12}. Low-frequency ($0.01$–$30$~Hz) quasi-periodic oscillations (LFQPOs) are a characteristic temporal feature 
commonly observed during the hard and intermediate spectral states of transient BHCs. Based on their properties (centroid frequency, quality 
factor or Q-value, rms amplitude, noise characteristics, time lag, etc.), LFQPOs are categorized into three types: A, B, and C \citep{Casella05}.
Typically, type-C QPOs are observed to evolve monotonically during the HS and HIMS states in both the rising and declining phases of an outburst 
\citep{Nandi12,D13}. In contrast, type-B and type-A QPOs are sporadically detected during the SIMS spectral state.
According to the shock oscillation model (SOM) introduced by Chakrabarti and his collaborators \citep{MSC96,RCM97}, LFQPOs originate 
due to oscillations of the shock. In the TCAF solution (Chakrabarti \& Titarchuk 1995), a hot Comptonizing region, known as `CENBOL', 
naturally forms during the accretion process in the post-shock region. In the SOM framework, shock oscillations occur due to the heating 
and cooling effects within the CENBOL. The model suggests that sharp type-C QPOs arise from resonance oscillations of the shock, while 
type-B QPOs occur either due to the non-satisfaction of the Rankine-Hugoniot condition or due to a weakly resonating CENBOL. The broader type-A 
QPOs are attributed to weak oscillations of the shockless centrifugal barrier \citep[see][]{C15}. To explain the evolution of QPO frequencies 
during the rising and declining harder spectral states of transient BHCs, a time-varying form of the SOM, namely the propagating oscillatory 
shock (POS) model, was introduced in 2005 \citep[see][]{C05,C08}.

The Galactic `faint' transient BHC MAXI~J1834$-$021 was first detected by MAXI/GSC \citep{Negoro23} on February 05, 2023 (MJD 59980). Unfortunately,
it was reported nearly a month later, on 2023 March 6 (MJD 60009). Based on the source flux observed from February 28, 2023 (MJD 60003.95) to March 1, 2023
(MJD 60004.53), \citet{Negoro23} estimated the source location as (RA, Dec)$ = (278^\circ.634, -2^\circ.130) = (18^{\rm h}34^{\rm m}32^{\rm s}, -02^\circ07'47'')$ (J2000)
with a 90\% confidence level. The maximum average flux of the source during the aforementioned period was observed to be
$18 \pm 4$~mCrab in the $4$–$10$ keV MAXI/GSC band, with a $1\sigma$ error. Subsequently, MAXI~J1834-021 was monitored in multiple wavelength bands,
including X-rays \citep[Swift, NICER, NuSTAR; see][]{Kennea23,Marino23,Homan23}, and optical \citep[LCO and Faulkes telescopes; see][]{Saikia23}.
The source was not detected in the $15.5$~GHz band by the AMI-LA radio telescope \citep{Bright23}. Based on preliminary spectral and temporal analyses,
including the detection of LFQPOs using NICER data, \citet{Homan23} confirmed the source as a black hole low-mass X-ray binary.

In \citet[][hereafter Paper-I]{DC25}, detailed spectral and temporal studies of MAXI J1834-021 has been done to understand accretion flow properties 
of the source during its entire 2023 active phase using NICER data. The study suggests that it showed double `failed' outbursts during its 2023 epoch. 
The spectral study using both phenomenological (combined disk blackbody plus powerlaw) and physical (TCAF) models show a strong dominance of nonthermal 
photons from `hot' Compton cloud throughout the 2023 active phase. During entire active phases, only harder spectral states are observed. 
The TCAF model also estimates mass of the source in the range of $12$–$12.8~M_\odot$. 
A monotonic evolution of the QPO frequency from higher to lower frequencies are observed, which is analogous to the QPO evolution in the declining phases 
of outbursts of a few transient BHCs (e.g., 2005 outburst of GRO J1655-40, 2010-11 outburst of GX 339-4, 2011 \& 2011 outbursts of H 1743-322).

Mass is a crucial intrinsic parameter in astronomical sources, particularly in compact objects. Accurately determining the mass of the central BH 
in a binary system is essential for understanding accretion-ejection processes around it. However, dynamical measurements of BH masses are sometimes 
not possible due to the lack of information about the binary companion. In such cases, alternative methods are employed to estimate the mass of BHs.  
Some of these alternative methods include: the QPO frequency ($\nu$)-photon index ($\Gamma$) correlation method \citep{ST07, ST09}; 
the high-frequency QPO (HFQPO)-spin ($a$) correlation method \citep{Motta14}; the inverse scaling relation with observed HFQPOs \citep{RM06};  
the QPO frequency evolution method \citep{Iyer15, D25}; and the TCAF model-based spectral fitting method \citep{Molla16}.  
The $\nu$-$\Gamma$ correlation method is widely used to estimate the masses of many neutron stars, black holes, and AGNs, while method based 
on TCAF model is also used to estimate the masses of more than $15$ stellar mass BHCs.  

Black hole spin is a fundamental parameter that influences both accretion and ejection processes. In X-ray binaries, spins are primarily estimated 
using two methods: the continuum-fitting method (CFM) and X-ray reflection spectroscopy (RS). Both techniques rely on fitting the observed X-ray 
spectra but are sensitive to different aspects of the accretion flow. The CFM (e.g. \texttt{kerrbb}) involves modeling the thermal X-ray emission from a geometrically 
thin, optically thick Keplerian accretion disk \citep[see, e.g.,][]{Zhang97,Shafee06,Gou10,Steiner11}, whereas the RS method models (e.g., 
\texttt{LAOR}, \texttt{relxill}) the relativistically broadened iron K$_\alpha$ line and the associated reflection continuum 
\citep{Laor91,Garcia14,Mondal16}.

In this study, we analyze the evolution of the QPO frequency during the declining phase of the primary outburst of MAXI~J1834-021 
with the propagating oscillatory shock (POS) model. We also performed spectral study using combined NICER plus NuSTAR data with both 
phenomenological and physical models. The POS model fit as well as spectral study with the \texttt{TCAF, kerrbb} models provides a good 
estimation of the mass of the BH. The \texttt{kerbb} model also estimates spin, distance, inclination angle of the source. 
The paper is organized as follows: \S 2 describes observations, data reduction, and analysis procedures. In \S 3, we present the results, 
while \S 4 discusses our findings and presents the conclusions.

\section{Observation and Data Analysis}

\subsection{Observations}

NICER started to monitor MAXI~J1834-021 roughly seven days after the announcement of the discovery by MAXI/GSC on March 06, 2023 (MJD 60009). 
The retrospective analysis of the MAXI/GSC suggests that the source was detected on February 05, 2023 (MJD 59980). In Paper-I, a detailed timing and 
spectral analysis has been made for $95$ observations of the NICER/XTI instrument from March 07, 2023 (MJD = 60010.01) to October 04, 2023 
(MJD = 60221.37). A monotonic evolution of the QPOs ($2.12$-$0.12$~Hz) are observed from March 09, 2023 (MJD = 60012) to April 28 (MJD = 60062.29). 
Here, we have adopted these data for further analysis using POS model. Additionally, we use one available observation from the NuSTAR/FPMA satellite 
instrument on March 10, 2023 (MJD = 60013.23) to perform a combined NICER plus NuSTAR spectral study over a broad energy band of $0.5$–$79$~keV.  

\subsection{Data Reduction}  
We follow the standard data reduction procedures for the NICER and NuSTAR satellites.  

\subsubsection{NICER}  
{\it NICER} contains primary scientific payload, namely X-ray Timing Instrument \citep[XTI;][]{Gendreau12}. It is an independent satellite, attached 
as an external payload to the International Space Station. Its operating energy range is $0.2$–$12$~keV with a time resolution of $\sim0.1$~${\mu}$s 
and a spectral resolution of $\sim85$~eV at 1~keV. For data analysis, we use the online platform SciServer\footnote{\url{https://www.sciserver.org}} 
with HEASARC's latest HEASoft package, version 6.34. The Level 1 data files are processed with the {\fontfamily{qcr}\selectfont nicerl2} script in 
the latest CALDB environment to obtain fully calibrated Level 2 event files, which are then screened for non-X-ray events or bad data times. Light 
curves with $1$~s and $0.01$~s time bins in the energy bands $0.5$–$3$~keV, $3$–$10$~keV, and $0.5$–$10$~keV are then extracted using the task 
{\fontfamily{qcr}\selectfont nicerl3-lc}. To obtain spectra in the default energy band, we use the task {\fontfamily{qcr}\selectfont nicerl3-spect} 
with {\fontfamily{qcr}\selectfont SCORPEAN} background model. 

\subsubsection{NuSTAR}  
{\it NuSTAR} archival data, downloaded from the web archive, are reduced using the latest {\it NuSTAR} data analysis software 
({\fontfamily{qcr}\selectfont NuSTARDAS}, version 2.1.4a). Cleaned event files are produced using the {\fontfamily{qcr}\selectfont nupipeline} task 
with the latest calibration files. With the {\fontfamily{qcr}\selectfont XSELECT} task, a circular region of $60''$ centered at the source coordinates 
is chosen as the source region, while a circular region of the same radius, away from the source location, is chosen as the background region.  
The {\fontfamily{qcr}\selectfont nuproduct} task is then used to extract the spectrum, ARF, and RMF files.  
The extracted spectra are subsequently rebinned to have at least $20$ counts per bin using the {\fontfamily{qcr}\selectfont GRPPHA} task.  

\subsection{Data Analysis}

We use the task {\fontfamily{qcr}\selectfont lcstats} on NICER $1$~s time-binned light curves to determine count rates in different energy bands.  
To study power density spectra (PDS), we analyze $0.01$~s time-binned light curves in the $0.5$–$10$~keV band. To determine the parameters (centroid 
frequency, full-width at half-maximum [FWHM], and power) of the QPOs, the PDS are fitted with a Lorentzian model using the {\fontfamily{qcr}\selectfont 
XRONOS} package of HEASoft.

The $0.5$–$10$~keV NICER spectra are fitted with a combination of thermal disk blackbody (DBB) and power-law (PL) models in  
{\fontfamily{qcr}\selectfont XSPEC} (Arnaud 1996). To account for interstellar absorption, we use the {\fontfamily{qcr}\selectfont TBabs}  
model with the hydrogen column density ($N_H$) parameter set as free. The {\fontfamily{qcr}\selectfont smedge} model, with an edge energy of  
$\sim0.81$~keV, is used to compensate for instrumental features in the NICER spectra. For fitting combined NICER plus NuSTAR data, we use both 
phenomenological (DBB+PL) and theoretical ({\fontfamily{qcr}\selectfont nthComp, TCAF, kerbb}) models.

\begin{figure}
  \vskip -0.2cm
  \centering
    \includegraphics[angle=0,width=9.0cm,keepaspectratio=true]{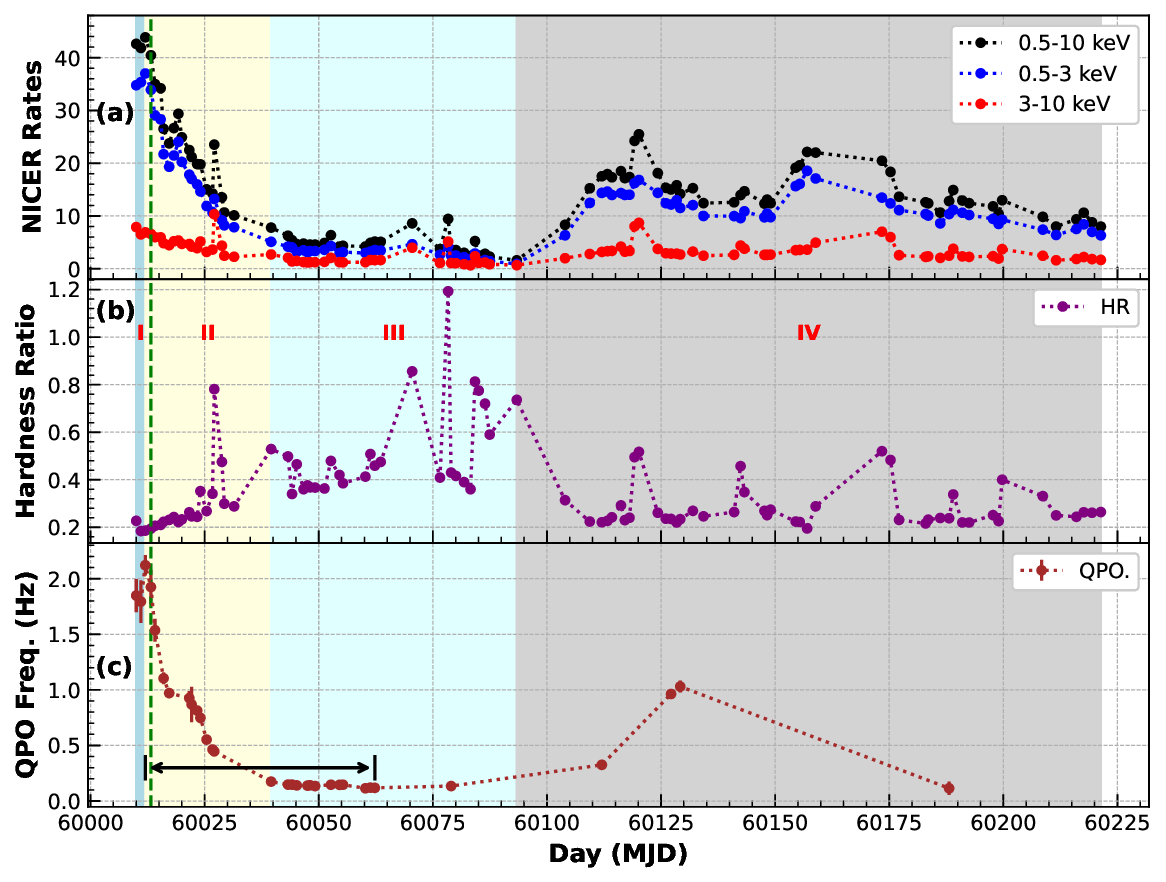}
\caption{(a) Variation of NICER count rates in the soft X-ray (SXR; $0.5$–$3$~keV), hard X-ray (HXR; $3$–$10$~keV),  
        and total X-ray (TXR; $0.5$–$10$~keV) energy bands, shown in the top two panels.  
        (b) Evolution of the hardness ratio (HR = HXR/SXR) is shown in the next panel.  
        (c) In the bottom panel, the variation of the observed QPOs is shown. The shaded regions mark different phases of the outburst profile.  
        The vertical green dashed-line corresponds to observation, whose detailed spectral study is made along with NUSTAR data. 
	The both-side arrow mark the duration of the QPO evolution whose detailed study made with the POS model.}
\label{lc-hr-qpo}
\end{figure}

\begin{figure}
  \vskip -0.4cm
  \centering
    \includegraphics[angle=0,width=8.8cm,keepaspectratio=true]{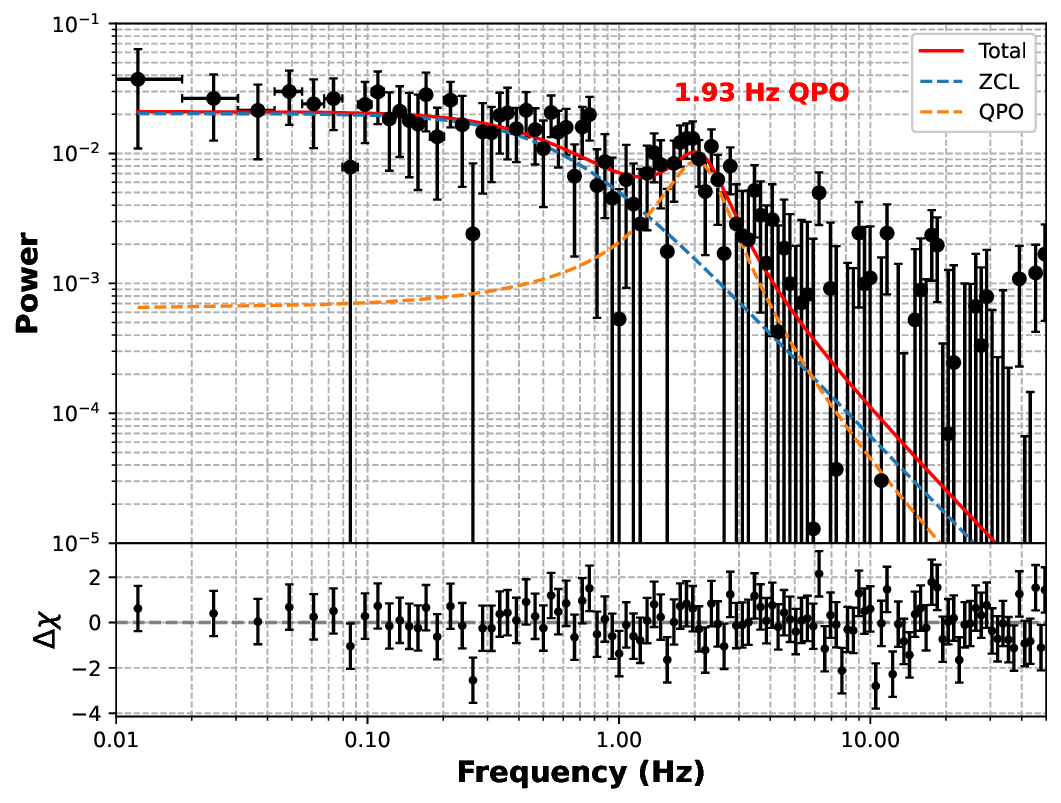}
    \caption{Fourier-transformed power density spectrum (PDS) of the $0.01$~s time-binned $0.5$–$10$~keV NICER light curve
        from March 10, 2023 (MJD 60013.23), fitted with combination of zero centric and QPO centric Lorentzian models. 
	The model fit reveals a prominent QPO at $1.93\pm0.16$~Hz.}
\label{pds}
\end{figure}

\begin{figure}
  \vskip -0.2cm
  \centering
    \includegraphics[angle=0,width=9.0cm,keepaspectratio=true]{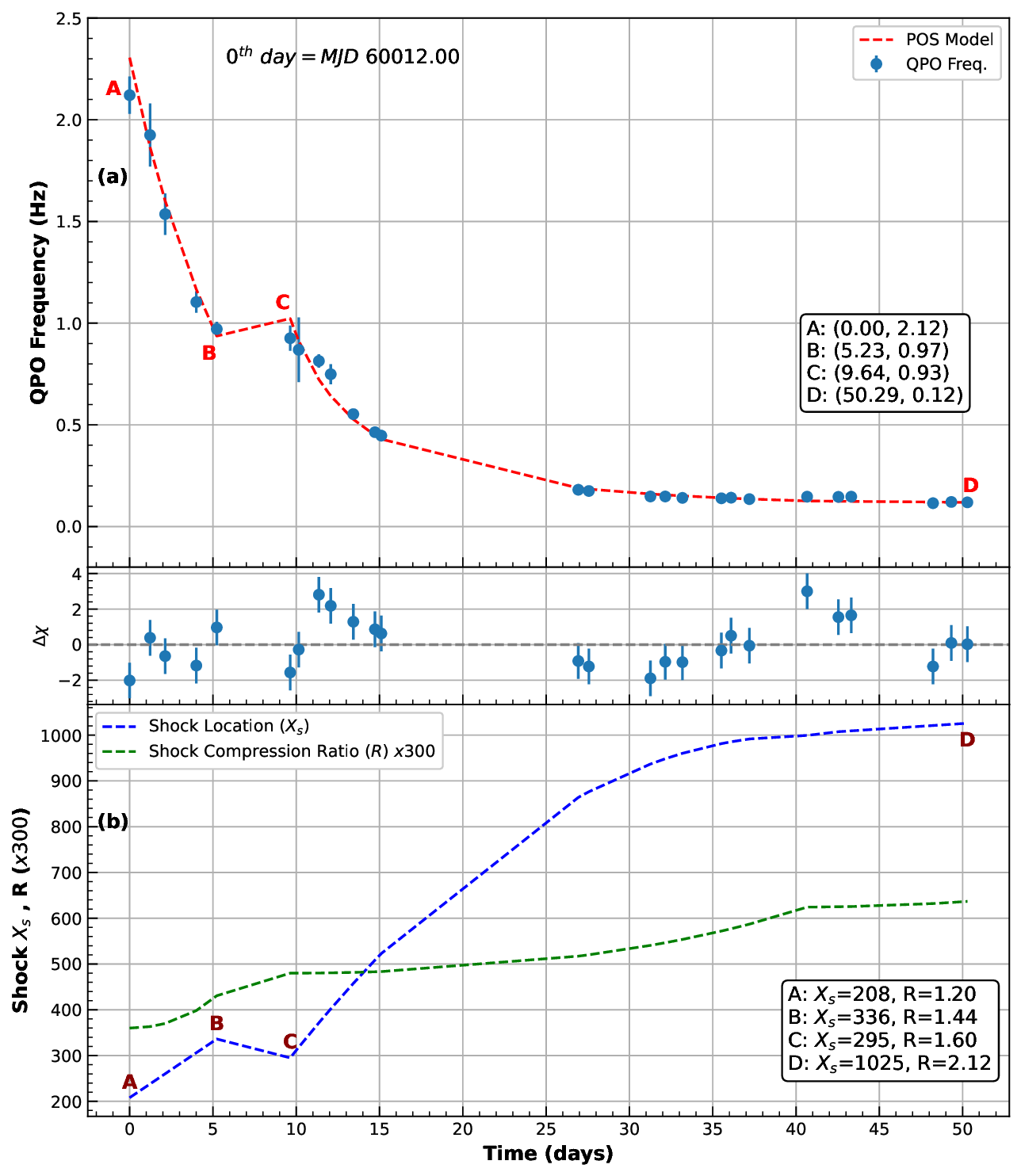}
        \caption{(a) Evolution of the observed QPO frequencies with time (in days) from March 9, 2023 (MJD 60012) to April 28, 2023 (MJD 60062.29), 
	fitted with the POS model (dashed red curve). (b) Variation of the POS model-fitted shock location ($X_s$ in Schwarzschild radius $r_s$) and  
        shock compression ratio (R) are shown in the bottom panel. Points A–D mark the start, end, or major transition phases of the evolution.}
\label{pos}
\end{figure}

\section{Results}

\subsection{Low frequency QPOs and Their Evolution}

In Paper-I, we studied detailed spectral and timing properties of MAXI~J1834-021 using archival data of NICER/XTI for $\sim 8$ months from 
March 07, 2023 (MJD = 60010.01) to October 04, 2023 (MJD = 60221.37). From the evolution of temporal and spectral properties, our studied 
period of the 2023 active phase had been sub-divided into four phases. The evolution of outburst profiles in hard and soft X-ray bands, 
hardness ratios (HRs), low frequency QPOs, and spectral model fitted parameters, fluxes, accretion flow parameters, etc. confirms double 
`failed' outbursts during 2023 active phase of the MAXI~J1834-021. The Phases I-III corresponds to the primary outburst and Phase-IV is 
classified as mini-outburst. Furthermore, spectral analysis with the phenomenological (combined disk black body plus powerlaw or with only 
powerlaw) and physical (TCAF) models, and nature of QPOs confirms existence of harder spectral states throughout the 2023 active phase. 
The Phases I-III are classified as HS (rising), HIMS (declining) and HS (declining) respectively, and Phase-IV as undivided harder spectral state.

Low-frequency QPOs (LFQPOs) are observed throughout the outbursts. Most importantly, in Phases II \& III, QPO frequencies are found to decrease 
monotonically from $2.12$~Hz to $0.19$~Hz within a period of $\sim 50$~days from March 09 to April 28, 2023 (MJD=60012-60062.29). This pattern of 
decreasing QPO frequencies is commonly observed in the declining harder states of transient BHCs \citep[see][]{C08,Nandi12,D13}. 
In Fig.~\ref{lc-hr-qpo}, we show variation of the soft X-ray (SXR; $0.5-3$~keV), hard X-ray (HXR; $0.5-3$~keV), total X-ray (TXR; $0.5-3$~keV), 
HR (ratio between HXR with SXR), QPO frequencies. These data are adopted from Paper-I. This evolution of the QPO frequency is further studied 
here with the POS model. In Fig.~\ref{pds}, we show Fourier transformed power density spectrum (PDS) from March 10, 2023 (MJD 60013.23), fitted 
with a combination of one zero-centered and one QPO-centric Lorentzian models. It detects a prominent signature of QPO at $1.93\pm0.16$~Hz.

\subsection{POS Model-Fitted QPO Evolution}

The propagating oscillatory shock (POS) model \citep{C05, C08} is the time-varying extension of the shock oscillation model \citep[SOM][]{MSC96, RCM97}, 
which explains the origin of QPOs due to the oscillation of the shock. Shock oscillations occur due to heating and cooling effects within the hot Compton 
cloud, or CENBOL. As a result, the oscillation of the shock corresponds to the oscillatory variation in the size of the Compton cloud.  

During the rising phase of transient BHCs, the oscillating shock moves inward due to an increased influx of Keplerian disk matter. Conversely, 
in the declining phase, the shock moves outward as the Keplerian disk matter supply decreases. According to the SOM, the QPO frequency 
($\nu_{\text{QPO}}$) follows an inverse relation with the shock location ($X_s$), given by: $\nu_{\text{QPO}} \sim X_s^{-3/2}$.
Thus, in the rising phase of an outburst, we generally observe increasing QPO frequencies, whereas in the declining phase, the opposite trend is seen.

The sharp Type-C QPOs are associated with the resonance oscillation of the shock, when matter cooling time scales roughly matches with the 
infall time ($t_{infall}$) scales. So, according to \citep{C08}, 
$$
\nu_{QPO} = \frac{\nu_{s0}}{t_{infall}}= \frac{c^3}{2GM_{BH}[R X_s (X_s-1)^{1/2}]}, 
\eqno{(1)}
$$
where $\nu_{s0} = c/r_s = c^3/2GM_{BH} = 10^5/(M_{BH}/M_{\odot})$ Hz is the inverse of the light crossing time of a black hole of mass $M_{BH}$
and $c$ is the velocity of light, $r_s = 2GM_{BH}/c^2$ is the Schwarzschild radius. Here $R (=\rho_+/\rho_-)$ is the shock compression 
ratio, i.e, ratio between the post- and pre-shock densities.

In the drifting or evolving shock scenario, $X_s = X_s(t)$ is time-dependent and can be expressed as
$$
X_s(t) = X_{s0} \pm \frac{v(t) t}{r_s},
\eqno{(2)}
$$
where `$-$' sign is used for inward and '+' sign is used for receding shock motion. The shock velocity $v(t)$ may be
accelerating, decelerating or constant and can be written as
$$
v(t) = v_0 \pm v_a t, 
\eqno{(3)}
$$
where $v_0$ is the initial shock velocity and $v_a$ is the velocity acceleration (`+')/deceleration (`-') term.

Generally during the rising phase of an outburst, the shock strength becomes weaker as the $R$ decreases. On the day
of the highest evolving QPO, it becomes even weaker, approaching $R \sim 1$. The opposite nature of $R$ can be seen 
in the declining phase of an outburst. The value of $R$ generally follows the equation
$$
\beta_s = \frac{1}{R} = \frac{1}{R_0} \pm \alpha t_d^2,
\eqno{(4)}
$$

where the `+' sign is used for the rising phase, and the `-' sign is used for the declining phase. Here, $R_0$ represents the initial 
compression ratio, and $\alpha$ is the controlling factor that determines the rate of increase or decrease of $R$ over time ($t_d$).

This model has been used to study the evolution of QPOs during the rising and/or declining phases of outbursts in several black hole 
candidates (BHCs), including GRO J1655-40 \citep{C05, C08}, GX 339-4 \citep{Nandi12}, H 1743-322 \citep{D13}, XTE J1550-564 
\citep{C09}, and Swift~J1727.8-1613 \citep{D25}. The model fit provides the instantaneous location, strength, and velocity of the oscillating 
shock. \citet{Iyer15}, \citet{Molla16} and \citet{D25} successfully applied this model to estimate the mass of BHCs IGR J17091-3624, MAXI~J1659-152, and Swift~J1727.8-1613, 
respectively. These studies motivated us to investigate the monotonic evolution of QPO frequencies during Phases II and III of the current 
outburst of MAXI J1834-021.

\begin{figure*}[!t]
\vskip 0.0cm
\centerline{
       \includegraphics[scale=0.6,angle=0,width=8.5truecm]{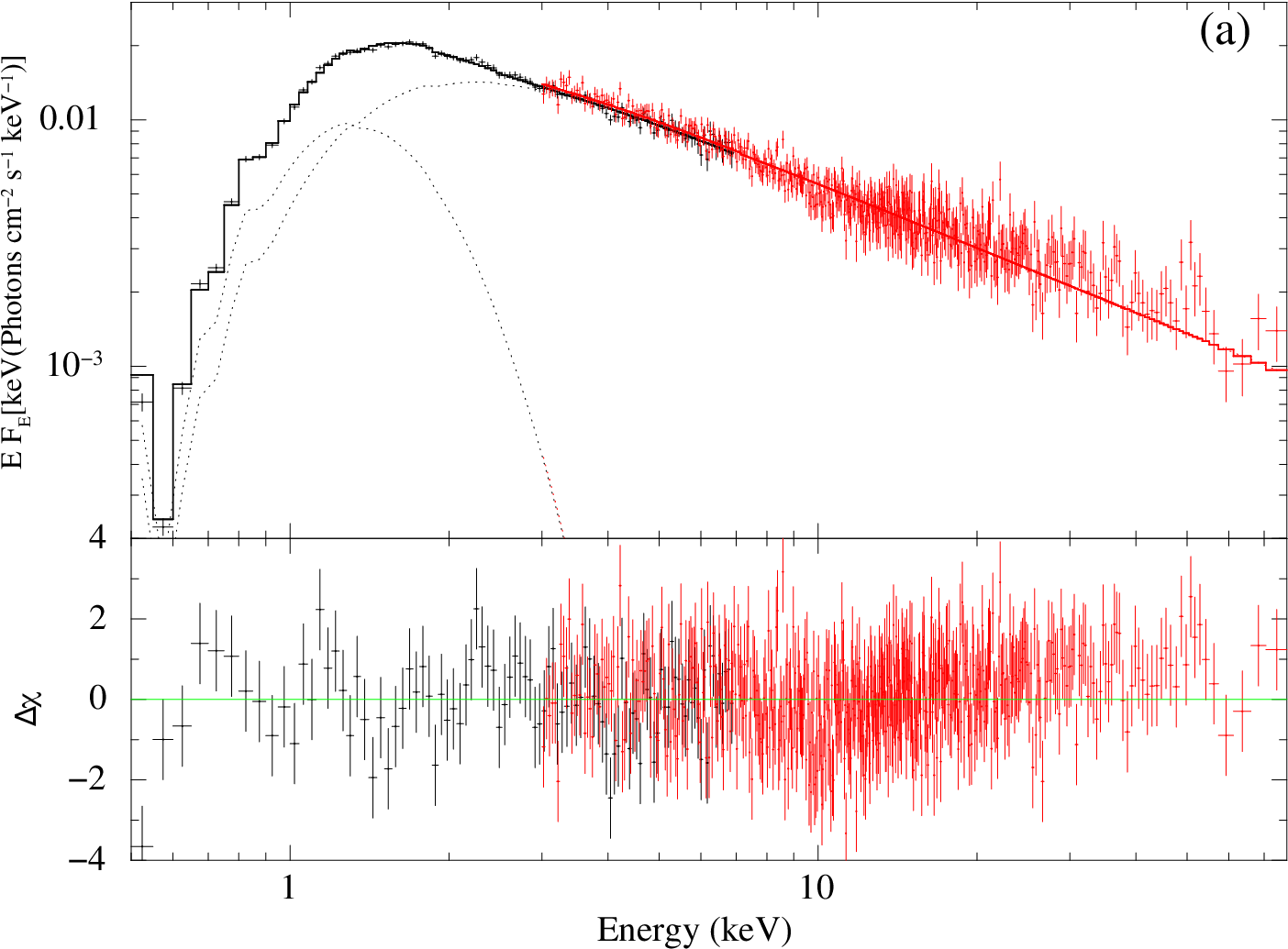}\hskip 0.3cm
       \includegraphics[scale=0.6,angle=0,width=8.5truecm]{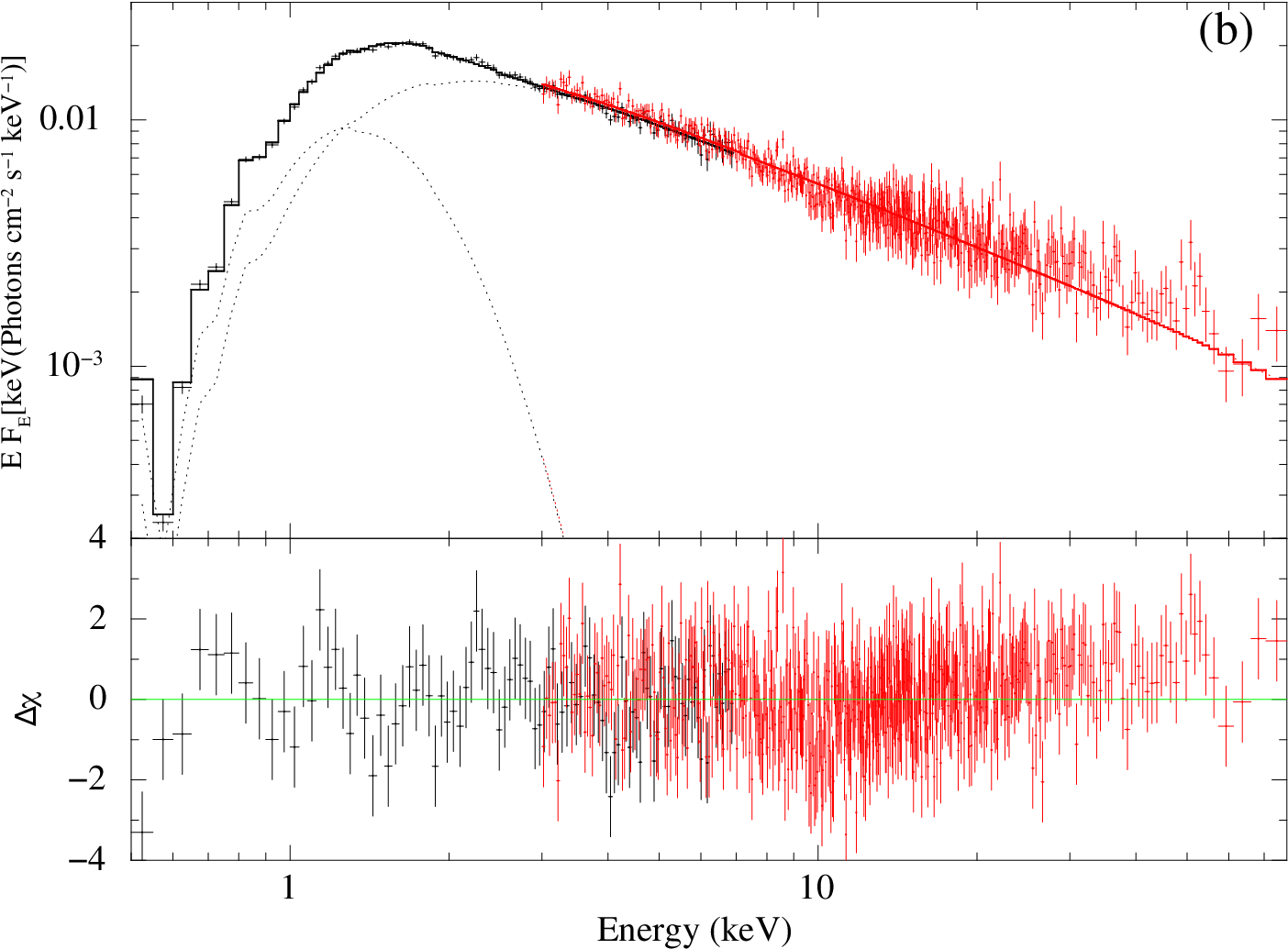}\\ }
\centerline{
       \includegraphics[scale=0.6,angle=0,width=8.5truecm]{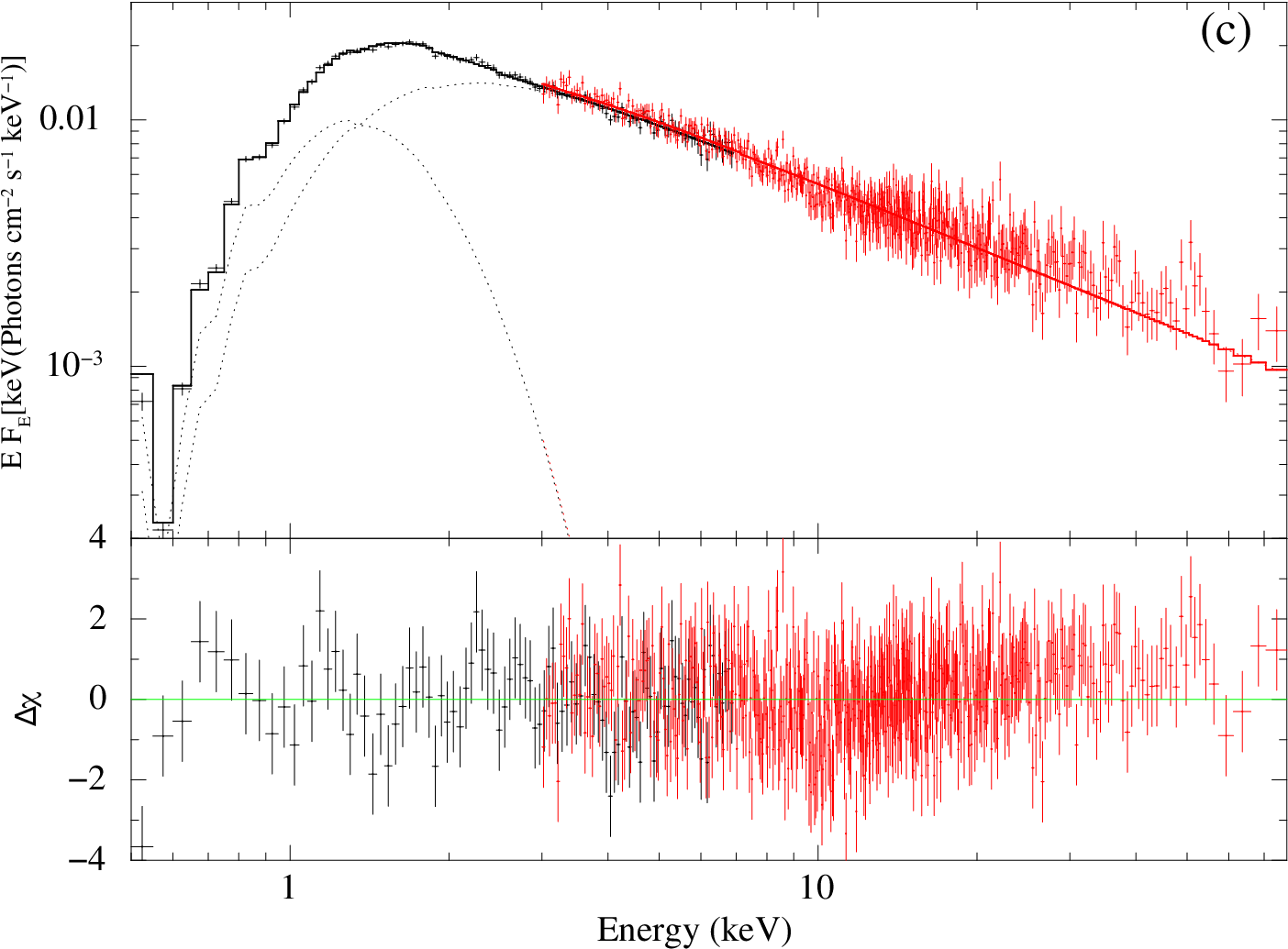}\hskip 0.3cm
       \includegraphics[scale=0.6,angle=0,width=8.5truecm]{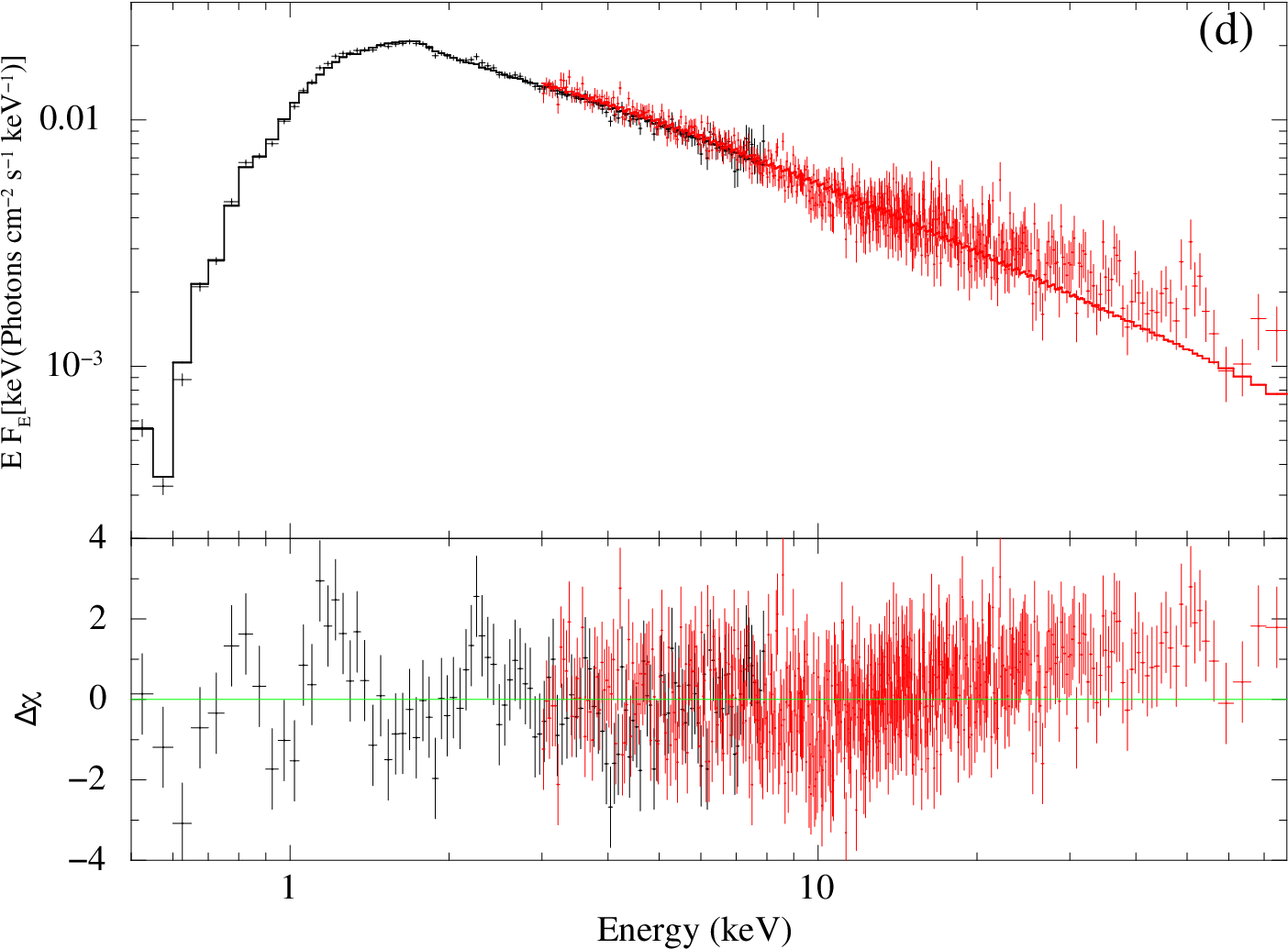}\\ }
        \caption{Combined NICER ($0.5-8$ keV) and NuSTAR ($3-79$ keV) spectrum of MAXI~J1834-021 on March 10, 2023 (MJD 60013.23), 
	fitted with different set of spectral models:  
        (a) $constant \otimes tbabs \otimes smedge (diskbb + powerlaw)$,  
        (b) $constant \otimes tbabs \otimes smedge (diskbb + nthComp)$,  
        (c) $constant \otimes tbabs \otimes smedge (kerrbb + powerlaw)$, and  
        (d) $constant \otimes tbabs \otimes smedge \otimes TCAF$.}
\label{spec}
\end{figure*}

After the initial rise in Phase-I, the QPO frequencies are found to monotonically decrease from $2.12$~Hz (2023 March 9; MJD = 60012) to $0.12$~Hz 
(2023 April 28; MJD = 60062.29). The monotonic evolution of the QPOs over this duration is quite similar to the evolution of QPOs during the 
declining phases of the 2005 outburst of GRO~J1655-40 \citep{C08}, the 2010-11 outburst of GX 339-4 \citep{Nandi12}, and the 2010 \& 2011 outbursts 
of H 1743-322 \citep{D13}. Similar to these earlier studies, the current evolution of the QPOs during the declining phase of the primary outburst is 
found to be well fitted with the POS model (see Fig.~\ref{pos}). The POS model-fitted QPO evolution, along with the variation of $\Delta \chi$, is shown in 
the top two panels of Fig.~\ref{pos}. The evolution of the POS model-fitted shock location ($X_s$) and compression ratio ($R$) is shown in the bottom panel 
of Fig.~\ref{pos}. 

During around $50$-day QPO evolution phase modeled by the POS model, the shock is found to move outward from $208~r_s$ to $1025~r_s$, with 
increasing shock strength. The compression ratio $R$ changes from $1.20$ to $2.12$. The start, end, and significant transition phases of the 
evolution are marked as points A–D. In the initial branch AB (MJD = 60012–60017.23), spanning $5.23$ days, a rapid decrease in the QPO frequency 
from $2.12$~Hz to $0.97$~Hz is observed. During this phase, $X_s$ and $R$ are found to vary from $208$–$336~r_s$, and $1.20$–$1.44$, respectively. 
After that, no QPOs are detected in branch BC (MJD = 60017.23–60021.63). Then, in branch CD (MJD = 60021.63–60062.29), spanning $\sim 41$ days, 
QPOs are found to decrease monotonically (from $0.93$ to $0.20$~Hz). During this phase, $X_s$ and $R$ vary from $295$–$1025~r_s$ and $1.6$–$2.12$, 
respectively. Except in the late evolution phase, the shock velocity is observed to vary between $10$–$20$~m/s. 

From Eq. (1), it is evident that the black hole mass ($M_{\rm BH}$) is an important parameter in the POS model. Thus, by fitting the POS model 
to the QPO frequency evolution, we are able to determine the mass of the newly discovered transient BHC MAXI~J1834-021. From the best POS fit, 
we estimate the black hole mass to be $M_{\rm BH} = 12.1 \pm 0.3~M_\odot$.

\subsection{Combined NICER plus NuSTAR Spectral Study}

To study the broadband nature of the source, we use the sole available NuSTAR observation along with the NICER observation from the same day. 
The combined NICER ($0.5$–$8$~keV) plus NuSTAR ($3$–$79$~keV) spectra from March 10 (MJD = 60013.23) are fitted with four sets of models:  
$(i)$ $constant\otimes tbabs\otimes smedge(diskbb + powerlaw)$, $(ii)$ $constant \otimes tbabs \otimes smedge (diskbb + nthComp)$,  
$(iii)$ $constant \otimes tbabs \otimes smedge (kerrbb + powerlaw)$, and $(iv)$ $constant \otimes tbabs \otimes smedge \otimes TCAF$ 
(see Fig.~\ref{spec}). The best-fit model parameters are listed in Table~\ref{table1}. Note, we have not fitted spectrum with the reflection 
based models (e.g., {\fontfamily{qcr}\selectfont pexrav, relxill}) as no prominent signature of the reflection feature were present in the data.

The model-fitted {\fontfamily{qcr}\selectfont TBabs} parameter $N_H$ and {\fontfamily{qcr}\selectfont smedge} parameter $edgeE$ are found to be 
consistent across all spectral fits. In sets $(i)$ and $(ii)$, the DBB temperatures $T_{\rm in}$ are found to be consistent, while in sets 
$(i)$–$(iii)$, the PL photon indices remain consistent in a moderate value of $1.87-1.88$. This values of $\Gamma$ are generally seen in HIMS of 
transient BHCs \citep{Nandi12, D13}. The sets $(i)$ and $(ii)$ confirms the presence of low temparature disk as $T_{\rm in}$ observed in 
between $0.30-0.35$~keV.

The physical {\fontfamily{qcr}\selectfont kerrbb} model includes four source parameters (spin, mass, inclination, and distance) and one flow 
parameter (disk rate). So, the best-fitted {\fontfamily{qcr}\selectfont kerrbb} model provides us four important intrinsic source parameters 
such as spin $a=0.13\pm0.09$, mass of the BH $M_{\rm BH} = 12.51\pm 1.44~M_\odot$, disk inclination $i = 80^\circ.4 \pm 2^\circ.08$, and 
distance $D = 9.43\pm 0.66$~kpc. 

Similarly, the physical {\fontfamily{qcr}\selectfont TCAF} model yields accretion flow parameters: Keplerian disk rate $\dot{m}_d = 0.036\pm 
0.002~\dot{M}_{\rm Edd}$, sub-Keplerian halo rate $\dot{m}_h = 0.589\pm 0.010~\dot{M}_{\rm Edd}$, shock location $X_s = 240\pm 4.2~r_s$, and  
shock compression ratio $R=1.13\pm 0.06$. Additionally, it provides the BH mass as $M_{\rm BH} = 12.41\pm0.13~M_\odot$. The presence of a higher 
halo accretion rate compared to the disk rate is consistent with the dominance of the PL flux over the DBB flux. The model-fitted $X_s$ and $R$ 
values are in agreement with the POS model-fitted QPO evolution values. Finally, the TCAF model-fitted $M_{\rm BH}$ is found to be consistent with 
the source mass values obtained from the POS and {\fontfamily{qcr}\selectfont kerrbb} models.

\begin{table*}
	\centering
	\caption{Spectral results of combined NICER (0.5-8\,keV) and {\it NuSTAR} (3-79\,keV) data on March 10, 2023}
	\label{tab:table2}
	\begin{tabular}{lcr||lcr} 
		\hline
  \hline
        \multicolumn{2}{c}{\textsc{Tbabs$\otimes$smedge(diskbb+powerlaw)}}&  ~      & \multicolumn{2}{c}{\textsc{Tbabs$\otimes$smedge(diskbb+nthComp)}} \\
   Tbabs  & $N_{\rm H}$ ($\times10^{22}~{\rm cm}^{-2}$) & 0.78$\pm$0.02   & Tbabs   & $N_{\rm H}$ ($\times10^{22}~{\rm cm}^{-2}$) & 0.72$\pm$0.02 \\
   diskbb & $kT_{\rm in}$ (keV) & 0.30$\pm$0.02                           & diskbb  & $kT_{\rm in}$ (keV) & 0.35$\pm$0.02 \\
	  & Norm        & 1180$\pm$32                                     &         & Norm  & 364$\pm$15            \\
 powerlaw & $\Gamma$      & 1.87$\pm$0.01                                 & nthComp & $\Gamma$      & 1.88$\pm$0.04 \\
 	  & Norm    & 0.04$\pm$0.001                                      &         & $kT_e$ (keV) & 999.5          \\
 	  &  $\chi^2/DOF$ & 617/578                                       &         & $kT_{bb}$      & 0.15         \\
F$_{DBB}$ & ($erg~cm^{-2}~s^{-1}$) & 1.75$\times10^{-11}$                 &         & Norm & 3.79                   \\  
F$_{PL}$  & ($erg~cm^{-2}~s^{-1}$) & 3.45$\times10^{-10}$                &         & $\chi^2/DOF$ & 620/576        \\
  	  & ~  & ~                                                        & F$_{DBB}$     & ($erg~cm^{-2}~s^{-1}$) &  1.70$\times10^{-11}$ \\
	  & ~  & ~                                                        & F$_{nthComp}$ & ($erg~cm^{-2}~s^{-1}$) &  3.42$\times10^{-10}$ \\
   \hline
\multicolumn{2}{c}{\textsc{Tbabs$\otimes$smedge(kerrbb + powerlaw)}}     &  ~    & \multicolumn{2}{c}{\textsc{Tbabs$\otimes$smedge$\otimes$TCAF}}\\
Tbabs    & $N_{\rm H}$ ($\times10^{22}~{\rm cm}^{-2}$) &   0.81$\pm$0.09 & Tbabs & $N_{\rm H}$ ($\times10^{22}~{\rm cm}^{-2}$) & 0.89$\pm$0.05 \\
kerrbb   & spin $a$  &        0.13$\pm$0.09                              &  TCAF & $\dot{m}_d$ ($\dot{M}_{Edd}$)  & 0.036$\pm$0.002  \\   
         & Inclination $i$ (deg) &     80.4$\pm$2.08                     &       & $\dot{m}_h$ ($\dot{M}_{Edd}$)  & 0.589$\pm$0.010  \\          
         & $M_{BH}$ ($M_\odot$)   &     12.51$\pm$1.44                   &       & $M_{BH}$ ($M_\odot$) & 12.41$\pm$0.13 \\         
         & $\dot{m}_d$ ($10^{18}$ g/s) &     0.09$\pm$0.02               & ~	& $X_s$ ($r_s$) & 240$\pm$4.2 \\        
         & Distance D (kpc)     &      9.43$\pm$0.66                     & ~     & $R$ & 1.13$\pm$0.06 \\
         & Norm                 &      1.14$\pm$0.16                     & ~     & Norm & 4.81$\pm$0.05 \\
powerlaw & $\Gamma$             &      1.87$\pm$0.01                     & ~     & $\chi^2/DOF$ & 695/587 \\
         & Norm                 &      0.04$\pm$0.001                    &  F$_{TCAF}$ & ($erg~cm^{-2}~s^{-1}$) & 3.47$\times10^{-10}$ \\
	 & $\chi^2/DOF$         &      615/574                           &  ~ & ~ & \\
F$_{kerrbb}$	& ($erg~cm^{-2}~s^{-1}$) &  1.83$\times10^{-11}$         &  ~ & ~ & \\ 
F$_{PL}$	& ($erg~cm^{-2}~s^{-1}$) &  3.04$\times10^{-10}$         &  ~ & ~ & \\ 
   \hline
      \end{tabular}
      \noindent{The {\fontfamily{qcr}\selectfont smedge} model parameter {\it edgeE} is found to be at $\sim 0.81$~keV for all four set of models.}\\
\label{table1}
\end{table*}

\subsection{MCMC Simulation and Contour Plots}

The combined NICER and NuSTAR broadband ($3$--$79$~keV) spectral study on March 10 (MJD = 60013.23) using the {\fontfamily{qcr}\selectfont kerbb} 
model allowed us to estimate four important source parameters: spin ($a$), mass of the black hole ($M_{\rm BH}$), disk inclination angle ($i$), 
and distance ($D$). To confirm these estimations, we performed a Markov Chain Monte Carlo (MCMC) simulation on the best-fitted spectrum in 
{\fontfamily{qcr}\selectfont XSPEC}\footnote{\url{https://heasarc.gsfc.nasa.gov/xanadu/xspec/manual/node43.html}}.
We employed the Goodman-Weare algorithm with $20$ walkers and a total chain length of $200{,}000$. The chain was initialized using a set of walkers 
drawn from a multi-dimensional Gaussian distribution with standard deviations $100$ times the parameter uncertainties and means equal to the best-fit 
model parameters (see Table~\ref{table1}). The simulated results were saved as a FITS file.

\begin{figure}
  \centering
    \includegraphics[angle=0,width=9.2cm,keepaspectratio=true]{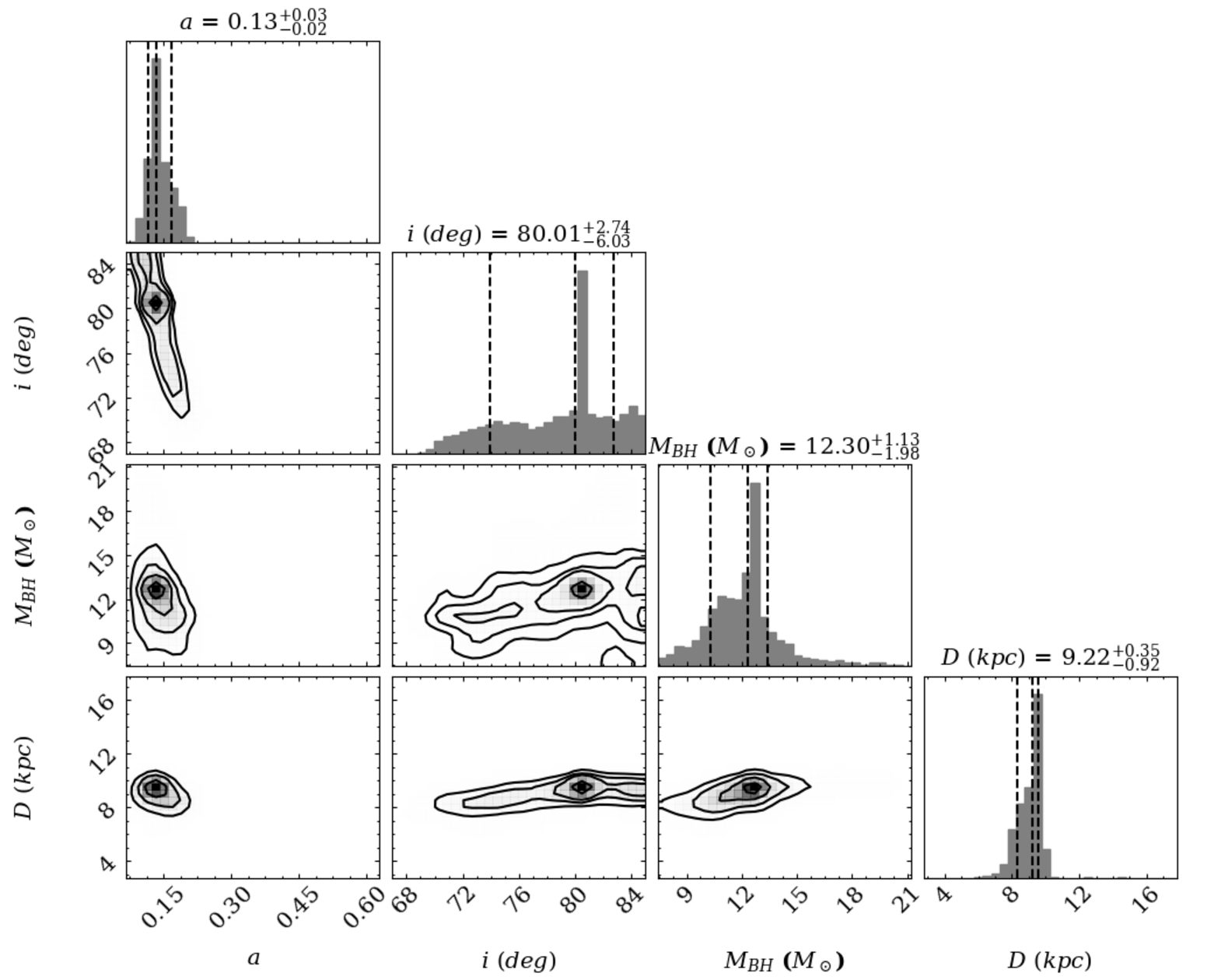}
	\caption{Contour plots ($1,2$, and $3\sigma$ levels) between \texttt{kerbb} model fitted intrinsic source parameters such as spin ($a$), 
	inclination angle ($i$), mass ($M_{BH}$), and distance ($D$). The Markov Chain Monte Carlo (MCMC) simulation generated fits file on best-fitted 
	spectrum (using combined model set $(iii)$: $constant \otimes tbabs \otimes smedge (kerrbb + powerlaw)$) has been used for the plots in pyXspecCorner 
	code of Federico Garcia.
	}
\label{contour}
\end{figure}

We then used the publicly available {\fontfamily{qcr}\selectfont pyXspecCorner}\footnote{\url{https://github.com/garciafederico/pyXspecCorner}} 
code by Federico Garcia to generate corner plots showing contours between different model parameters using the MCMC-generated FITS file. 
In Fig.~\ref{contour}, we present the $1\sigma$, $2\sigma$, and $3\sigma$ confidence contours between the four intrinsic source parameters. 
Closed contours are observed at all three confidence levels for the following parameter pairs: $a$--$M_{\rm BH}$, $a$--$D$, and $D$--$M_{\rm BH}$. 
However, the $2\sigma$ and $3\sigma$ contours involving the inclination angle are not well closed. So, we can infer that within the $1\sigma$ 
confidence level, all four intrinsic parameters are reasonably well constrained.

In Fig.~\ref{contour}, the MCMC-derived best-fit model parameters along with their uncertainties are displayed. The intrinsic source parameters 
obtained from this method are: spin $a = 0.13^{+0.03}_{-0.02}$, disk inclination $i = 80^\circ.0^{+2.7}_{-6.0}$, black hole mass 
$M_{\rm BH} = 12.3^{+1.1}_{-2.0}~M_\odot$, and distance $D = 9.2^{+0.4}_{-0.9}$~kpc.

In Paper~I, from a detailed spectral study of the outburst using the TCAF model, the black hole mass $M_{\rm BH}$ was estimated to lie within the 
range $12$--$12.8~M_\odot$, with an average of $12.3~M_\odot$, i.e., $12.3^{+0.5}_{-0.3}~M_\odot$. Here, the mass estimates obtained using the POS, 
{\fontfamily{qcr}\selectfont kerbb}, and TCAF models are $12.1 \pm 0.3~M_\odot$, $12.3^{+1.1}_{-2.0}~M_\odot$, and $12.4 \pm 0.1~M_\odot$, 
respectively. By combining these methods, we estimate the probable mass of MAXI~J1834--021 to be $12.3^{+1.1}_{-2.0}~M_\odot$.

\subsection{Origin of the Observed QPO}

The QPO frequency observed on the day of the combined NICER plus NuSTAR spectral analysis was $1.93 \pm 0.16$~Hz (see Fig.~\ref{pds}). The Lorentzian 
model fit yielded a line width (FWHM) of $0.538 \pm 0.122$ and a line normalization (power) of $0.0096 \pm 0.0011$. The corresponding quality factor 
($Q = \nu_{\rm QPO}/{\rm FWHM}$) and amplitude (in \% rms, where ${\rm rms} = (\pi/2) \times {\rm Power} \times {\rm FWHM}$) were found to be $3.58$ 
and $9.02$, respectively.

Assuming that type-C QPOs originate from the oscillation of a shock, \citet{D14} calculated the QPO frequencies for three BHCs using shock parameters 
obtained from TCAF model spectral fits. The close agreement between the theoretically predicted and the observed QPO frequencies strongly supported 
the hypothesis that shock oscillation was indeed responsible for the origin of the QPOs. This approach effectively predicts a temporal feature from 
spectral parameters. The same shock parameters---shock location ($X_s$) and compression ratio ($R$)---used to characterize the size of the Compton 
cloud and the matter densities in the pre-shock and post-shock regions are also used to calculate QPO properties via the POS model. Using the TCAF 
model-fitted values of $M_{\rm BH}$, $X_s$, and $R$ in Eq.~(1), we obtain a theoretical QPO frequency of $1.92 \pm 0.16$~Hz. The consistency between 
this calculated and observed $\nu_{\rm QPO}$ further confirms shock oscillation as the likely origin of the detected QPO in our studied observation 
of MAXI~J1834-021.

\section{Discussion and Concluding Remarks}

The Galactic transient BHC MAXI J1834-021 was detected on February 05, 2023 (MJD = 59980) by MAXI/GSC and it X-ray activity was continued for the 
next $\sim 10$ months. Due to almost one month delay in the report of the discovery (March 06, 2023), we have missed initial rising phase data of 
the source. In Paper-I, we studied the spectral and timing properties of the source during the initial $\sim 8$ months (2023 March 7 to October 4; 
MJD = 60010.01-60221.37) to understand accretion flow dynamics of the source during its 2023 active phase. From the the daily variations in X-ray 
intensities in the soft ($0.5$–$3$~keV, SXR), hard ($3$–$10$~keV, HXR), and total ($0.5$–$10$~keV, TXR) energy bands, nature of the osberved QPOs, 
and spectral fitted parameters, fluxes, etc. reveals four distinct stages during the outburst (see shaded regions of Fig.~\ref{lc-hr-qpo}). 
The Phases I-III are classified as primary and Phase-IV is classified as mini-outburst. Spectral studies were made with both phenomenological (combined 
disk blackboody plus powerlaw or only powerlaw) and physical ({\fontfamily{qcr}\selectfont TCAF}) models. The {\fontfamily{qcr}\selectfont TCAF} model 
also estimated mass of the source.

In Phases II \& III of the outburst i.e., in the declining phase of the primary outburst, a monotonic evolution of the QPO frequency from $2.12$ to 
$0.20$~Hz (in between MJD = 60012–60062.29) is observed. Here, To understand the nature of this QPO evolution from a physical point of view, we studied 
it using the POS model (see Fig.~\ref{pos}). The POS model fit allowed us to determine the physical flow parameters such as instantaneous location of 
the shock, compression ratio, velocity, etc. The shock is found to recede from $208$–$1025~r_s$ with a slowly increasing strength. The best-fitted 
POS model also enabled us to estimate the mass of the BH, as $M_{BH}$ is an essential parameter of the model (see Eq. 1). The mass of the BH is 
obtained as $M_{BH} = 12.1\pm0.3~M_\odot$. This is consistent with the mass range ($12-12.8~M_\odot$) of the source reported in Paper-I.

The spectral study throughout the 2023 active phase was done using $0.5-10$~keV NICER data in Paper-I. The study confirms that the presence of harder 
(HS and HIMS) spectral stattes with a dominance of nonthermal fluxes from `hot' Compton cloud i.e., sub-Keplerian halo accretion rate over the disk rate. 
Here, we make an attempt to study broadband nature (in $0.5-79$~keV) of the source using combined NICER ($0.5-10$~keV) plus NuSTAR ($3-79$~keV) data 
from March 10, 2023 (see, Fig.~\ref{spec} and Table~\ref{table1}). The spectrum has been studied with four different set of phenomenological as well 
as physical models. The combined DBB plus PL, and combined DBB plus nthComp infers presence of cooler disk with temparature $T_{in}$ in between 
$0.30-0.35$~keV. The TCAF model fit also tells about the lower Keplerian disk accretion rate which is supposed to the origin of the DBB photons, 
comprared to the sub-Kelerian halo rate. The kerrbb model also finds lower disk rate. The photon indices obtained from DBB+PL, DBB+nthComp, and kerbb+PL 
model fits are found to be consistent with values in a constant range of $1.87-1.88$. This $\Gamma$ values are generally seen in HIMS of tranisent 
stellar mass BHCs. 

The broadband analysis using the physical {\fontfamily{qcr}\selectfont TCAF} and {\fontfamily{qcr}\selectfont kerrbb} models allowed us to estimate 
four intrinsic source parameters: mass, spin, disk inclination angle, and distance. To validate the intrinsic parameters obtained from the 
{\fontfamily{qcr}\selectfont kerrbb} model, we performed an MCMC simulation in {\fontfamily{qcr}\selectfont XSPEC}. Confidence contours were plotted 
using the MCMC-generated fits file and the publicly available {\fontfamily{qcr}\selectfont pyXspecCorner} code. These contours provided constraints 
on the fitted parameter values and their uncertainties. From this method, we estimated the spin $a = 0.13^{+0.03}_{-0.02}$, disk inclination 
$i = 80^\circ.0^{+2.7}_{-6.0}$, mass $M_{\rm BH} = 12.3^{+1.1}_{-2.0}~M_\odot$, and distance $D = 9.2^{+0.4}_{-0.9}$~kpc.

Combining the $M_{\rm BH}$ estimates from {\fontfamily{qcr}\selectfont TCAF} model fits in Paper-I ($12$–$12.8~M_\odot$, or $12.3^{+0.5}_{-0.3}~M_\odot$), 
the present TCAF fit ($12.41 \pm 0.13~M_\odot$), the {\fontfamily{qcr}\selectfont POS} model fit ($12.1 \pm 0.3~M_\odot$), and the above MCMC-simulated 
{\fontfamily{qcr}\selectfont kerrbb} results, we infer the probable mass of MAXI~J1834$-$021 as $12.3^{+1.1}_{-2.0}~M_\odot$. 
The estimated low spin ($a = 0.13^{+0.03}_{-0.02}$) places MAXI~J1834$-$021 among a few sources with spins below 0.3, such as A0620$-$00 
($a = 0.12 \pm 0.19$, \citealt{Gou10}), H1743$-$322 ($a = 0.2 \pm 0.3$, \citealt{Steiner11}), and LMC~X-3 ($a = 0.25^{+0.20}_{-0.29}$, \citealt{Steiner14}). 
The high inclination angle ($i = 80^\circ.0^{+2.74}_{-6.03}$) makes this source a suitable candidate for soft-lag studies (see, e.g., \citealt{Dutta16}). 
A detailed study of the lag properties will be presented elsewhere. 

A prominent QPO of $1.93\pm0.16$~Hz is observed in NICER data on March 10, 2023, whose detailed spectral study has been done in this paper (see inset 
in Fig.~\ref{lc-hr-qpo}). To find origin of this QPO, we followed method as described in \citet{D14}. The frequency of the primary QPO is being 
calculated from the TCAF model fitted shock parameters ($X_s$ and $R$). If the shcok oscillation being the origin of the observed QPO, the calculated 
QPO should match with the observed one. Here, using TCAF model fitted $X_s$, $R$, and $M_{BH}$ in Eq. (1), we calculated theoretical $\nu_{ThQPO}$ as 
$1.92\pm0.16$, which matches with the obsered $\nu_{QPO}$. This actually confirm oscillation of the shock as the origin of the observed QPO on March 10, 2023.

A brief summary of our findings in this {\it paper} is as follows:

\begin{enumerate}[i)]
    \item The entire period of the 2023 activity of MAXI~J1834-021 exhibits variations in soft (SXR; 0.5–3~keV), hard (HXR; 3–10~keV), and total (TXR; 
	  0.5–10~keV) X-ray count rates, QPO frequencies, etc., which classify the event into four phases of a double outburst.
    \item Strong signatures of LFQPOs are observed throughout the outburst. The monotonic evolution of QPOs during the declining phase of the primary outburst 
	  is further studied with the POS model. A receding shock with slowly increasing strength is found. The POS model also estimated the mass of the BH 
	  as $12.1 \pm 0.3~M_\odot$.
    \item The combined NICER plus NuSTAR spectral analysis on 2023 March 10, in a broad energy band with four sets of models not only allowed us to 
	  understand the nature of the source from a physical perspective but also estimates the intrinsic source parameters. The source was found in 
	  hard-intermeidate spectral state with higher dominance of non-thermal photons.
    \item The spectral fit with the physical {\fontfamily{qcr}\selectfont kerrbb} model estimates the mass, spin, distance, and inclination angle as 
	  $M_{\rm BH} = 12.3^{+1.1}_{-2.0}~M_\odot$, $a = 0.13^{+0.03}_{-0.02}$, $D = 9.2^{+0.4}_{-0.9}$~kpc, and $i = 80^\circ.0^{+2.7}_{-6.0}$, 
	  respectively.
    \item The spectral fit with the {\fontfamily{qcr}\selectfont TCAF} model finds a higher halo accretion rate ($\dot{m}_h$) compared to the disk 
	  rate ($\dot{m}_d$). The shock parameters ($X_s = 240 \pm 4.2$ and $R = 1.13 \pm 0.06$) signify the presence of a large Compton cloud with 
	  moderate strength. The model also estimates the mass of the source as $12.4 \pm 0.1~M_\odot$.
    \item Combining the estimated $M_{\rm BH}$ values from the {\fontfamily{qcr}\selectfont POS}, {\fontfamily{qcr}\selectfont kerrbb}, and 
	  {\fontfamily{qcr}\selectfont TCAF} models, we predict the most probable mass of MAXI~J1834--021 as $12.3^{+1.1}_{-2.0}~M_\odot$, 
	  or within the range of $11.2$--$14.3~M_\odot$.
    \item The match between the theoretically estimated QPO frequency from the {\fontfamily{qcr}\selectfont TCAF} model-fitted shock parameters and 
	  the observed one confirms shock oscillation as the origin of the prominent $1.93 \pm 0.16$~Hz QPO on March 10, 2023.
\end{enumerate}

\section*{Acknowledgements}

This work made use of NICER/XTI and NuSTAR/FPMA data supplied by the High Energy Astrophysics Science Archive Research Center (HEASARC) archive.
D.D. acknowledge the visiting research grant of National Tsing Hua University, Taiwan (NSTC NSTC 113-2811-M-007-010). 
H.-K. C. is supported by NSTC of Taiwan under grant NSTC 113-2112-M-007-020.

\end{document}